\documentstyle[eqsecnum, epsf, aps]{revtex}

\begin{document}

\draft

\title{ On Gravity and the Uncertainty Principle}

\author{ Ronald J. Adler\thanks{adler@relgyro.stanford.edu}} 
\author{ David I. Santiago\thanks{david@spacetime.stanford.edu}}
\address{ Gravity Probe B, W.W.Hansen Experimental Physics Laboratory, and 
Department of Physics, Stanford University, Stanford, Ca 94305-4085}

\date{\today}

\maketitle

\begin{abstract}
Heisenberg showed in the early days of quantum theory that the uncertainty
principle follows as a direct consequence of the quantization of
electromagnetic radiation in the form of photons. As we show here the
gravitational interaction of the photon and the particle being observed
modifies the uncertainty principle with an additional term. From the
modified or gravitational uncertainty principle it follows that there is an
absolute minimum uncertainty in the position of any particle, of order of
the Planck length. A modified uncertainty relation of this form is a
standard result of superstring theory, but the derivation given here is
based on simpler and rather general considerations with either Newtonian
gravitational theory or general relativity theory.
\end{abstract}


\section{Introduction}

Max Planck discovered the eponymous constant $\hbar$ when studying black body
radiation in 1900 \cite{weav}. He realized immediately that the constants
$\hbar$ and $c$ and  $G$ determine a natural scale, now called the Planck 
scale, which is easily gotten by dimensional analysis \cite{mtw}. The Planck
distance, time, mass, and energy are
\begin{eqnarray}
L_p \equiv \sqrt{\frac{G \hbar}{c^3}} \simeq 1.6 \times 10^{-35}\text{ m} \, , 
\ \ \ T_p \equiv \frac{L_p}{c} = \sqrt{\frac{G \hbar}{c^5}} \simeq 0.54 \times
10^{-43}\text{ sec} \nonumber \\
M_p \equiv \frac{\hbar}{c L_p} = \sqrt{\frac{\hbar c}{G}} \simeq 2.2 \times 
10^{-8} \text{ kg}\, , \ \ \ E_p \equiv \sqrt{\frac{\hbar c^5}{G}} \simeq 
2.0 \times 10^9 \text{ J} = 1.2 \times 10^{19} \text{ GeV} \, .\label{planck}
\end{eqnarray}
From its construction the Planck scale should be relevant when the system
considered is quantum mechanical ($\hbar$), involves high velocities and high
energies ($c$), and gravity is important ($G$). One such system is the very
early universe. Another is the collision of elementary particles such as
quarks at about the Planck energy, but to achieve the Planck energy in a
laboratory would probably require an accelerator about the size of a
galaxy. Yet another system in which the Planck scale is relevant is the
cloud of virtual particles surrounding any real particle, since the virtual
particles may in principle have arbitrarily high energies.

Much work has gone into constructing a quantum theory of gravity
appropriate to the Planck scale, but with little practical success. The
only theory thus far that seems to be a plausible candidate is superstring
theory \cite{str}.

Motivation for the present work originated at a talk by John Schwarz at
the Stanford Linear Accelerator Center in 1996, in which he presented a
modified uncertainty principle as a result of superstring theory and scale
inversion symmetry. He asked if, in view of its simplicity, it might be
more general than superstring theory or any particular quantum gravity
theory, and perhaps derivable by simpler means. We hope this work partially
answers that question.

We do not consider any specific quantum gravity theories here, but instead
show that the Heisenberg uncertainty principle is modified when we combine
quantum theory and some basic concepts of gravity. We give four separate
derivations of the modified uncertainty principle: dimensional analysis in
Newtonian theory, an approximate calculation in Newtonian theory,
dimensional analysis in general relativity theory, and an approximate
calculation in general relativity theory \cite{oha}. All four
derivations are heuristic and somewhat rough, as befits a discussion of the
uncertainty principle. Moreover the derivations based on Newtonian theory
should not be taken too seriously since the Newtonian theory is
action-at-a-distance, which is certainly inappropriate for a photon moving
at $c$; our purpose in including the Newtonian derivation is to show that the
modified uncertainty principle appears to follow from rather general
considerations on gravity, in particular that the additional gravitational
term is linear in the energy or momentum of the photon.

A minimum position uncertainty arises immediately from the modified
uncertainty principle, of order of the Planck distance. One might consider
this to be a minimum physically meaningful distance, and thereby question
whether any theory based on a a smaller scale, eg. a spacetime continuum,
really makes operational sense. Additionally, since all measurements
dealing with small distances in particle physics are really large momentum
scattering experiments, the very concept of a spacetime continuum is doubly
suspect. Further speculations of this nature are contained in the
conclusions.

\section{The Uncertainty Principle a la Heisenberg}

Heisenberg in 1923 obtained the uncertainty principle on very general
grounds, based only on general principles of optics and the quantization of
electromagnetic radiation in the form of photons \cite{sch,weav}. We recall his
approach briefly. Consider a wave scattering from an
electron into a microscope and thereby giving a measurement of the position
of the electron. According to optics and intuition, with an
electromagnetic wave of wavelength $\lambda$ we cannot obtain better precision 
than
\begin{equation}
\Delta x_H \approx \lambda \, . \label{lamb}
\end{equation}
Such a wave is quantized in the form of photons, each with a momentum
\begin{equation}
p=\frac{h}{\lambda} \, . \label{p}
\end{equation}
In order to interact with the electron an entire photon in the wave must
scatter and thereby impart to the electron a significant part of its
momentum, which produces an uncertainty in the electron momentum of about
$\Delta p \approx p$. Thus we obtain the standard Heisenberg position-momentum 
uncertainty relation
\begin{equation}
\Delta x_H \Delta p \approx \lambda\left( \frac{h}{\lambda} \right) \approx h
\approx \hbar \, , \label{up}
\end{equation}
No mention has been made here of the gravitational interaction between the
photon and the electron, which we consider below.

\section{Newtonian Theory, Dimensional Estimate}
	
We first estimate the effects of gravity in a very rough and heuristic way
using Newtonian gravitational theory, with the assumption that the photon
behaves as a classical particle with an effective mass equal to its energy
divided by $c^2$ \cite{oha}. Suppose the electron is in an experimental
region of characteristic size $L$, inside of which it interacts with the
photon. It will experience an acceleration due to gravity,
\begin{equation}
\ddot{\vec{r}} = - \frac{G (E /c^2)}{r^2} \hat{r} \, , \label{acc}
\end{equation}
where $r$ is the distance between electron and photon. During the
interaction, which occurs in characteristic time $L/c$, the electron will
acquire, due to gravity, a velocity and move a distance, given respectively
by
\begin{equation}
\Delta v \approx \frac{GE}{c^2 r^2} \left( \frac{L}{c} \right) \, , \ \ \ \
\Delta x_G \approx \frac{GE}{c^2 r^2} \left( \frac{L}{c} \right)^2 \, .
\end{equation}
These will be uncertain since the photon scatters electromagnetically from
the electron at some indeterminate time during the interaction. The
electron may be anywhere in the interaction region so the electron-photon
distance should be of order $r \approx L$, which is the only distance scale in
the problem. Since the photon energy is related to the momentum by $E=pc$ we 
may also express this as
\begin{equation}
\Delta x_G \approx \frac{Gp}{c^3} \, . \label{xG}
\end{equation}
Noting that the electron momentum uncertainty must be of order of the
photon momentum, and using the Planck length $L_{p}^{2} \equiv G \hbar/ c^3$
as a parameter, we have
\begin{equation}
\Delta x_G \approx \frac{G \Delta p}{c^3} = \left( \frac{G \hbar}{c^3}\right)
\frac{\Delta p}{\hbar}= L_{p}^{2} \frac{\Delta p}{\hbar} \, .\label{xG2}
\end{equation}
This is our main result. We add this uncertainty to the Heisenberg relation
(3) to obtain the modified uncertainty relation
\begin{equation}
\Delta x \approx  \frac{\hbar}{\Delta p}+ L_{p}^{2} \frac{\Delta p}{\hbar} \, .
\label{gup}
\end{equation}
We refer to this as the extended uncertainty principle - or more
descriptively as the gravitational uncertainty principle (GUP). Note that
it is invariant under
\begin{equation}
\frac{\Delta p L_p}{\hbar} \longleftrightarrow \frac{\hbar}{\Delta p L_p} \, .
\label{inv} 
\end{equation}
That is, it has a kind of momentum inversion symmetry \cite{str}.

\section{Newtonian Theory, Approximate Calculation}

We may also make a more explicit estimate than the above using Newtonian
theory.  As before we suppose the electron is in an experimental region of
characteristic size $L$ and interacts with the photon as it crosses the
region. The photon scatters electromagnetically from the electron at some
uncertain time and at some uncertain position inside the experimental
region. Consider first the transverse impulse, that is the motion imparted
to the electron perpendicular to the photon direction. We take the photon
direction to be $x$ and the transverse direction to be $y$ as in figure 2. 
The photon passes very rapidly so the electron moves very little in the time it
takes the photon to cross the experimental region, and we may thus take $y 
\approx y_0$ that is we use an impulse approximation. The acceleration is then
\begin{equation}
\ddot{y}=\frac{G(E/c^2)}{r^2} \left( \frac{y}{r} \right)= \frac{G (p/c)}{r^2} 
\left( \frac{y}{r} \right) \approx \frac{G(p/c)y_0}{(y_{0}^{2} + c^2 
t^2)^{3/2}} \, . \label{ay}
\end{equation}
We integrate this to get the transverse velocity impulse,
\begin{equation}
\Delta \dot{y} \approx \frac{2Gp}{c^2 y_0}\left( \frac{cT}{\sqrt{y_{0}^{2} 
+ c^2 T^2}}\right) \, ,\label{vy}
\end{equation}
where $T=L/c \ge y_0$ is the characteristic interaction time. We thus have 
roughly
\begin{equation}
\Delta \dot{y} \approx \frac{2Gp}{c^2 y_0} \, .\label{vy2}
\end{equation}
Due to this gravitational velocity impulse there will be a change in the
position of the electron, which is intrinsically uncertain, given
approximately by $\Delta y \approx (\Delta \dot{y} / 2) /T$  or
\begin{equation}
\Delta y_G \approx \frac{GpT}{c^2 y_0} \approx \frac{Gp}{c^3} \, ,\label{yG}
\end{equation}
where we have taken $y_0$ to be of order of but less than $L=cT$. This is the 
same result (\ref{xG}) as we obtained by somewhat more crude dimensional
arguments in section III.

A similar analysis can be done for motion in the longitudinal direction,
with the result
\begin{equation}
\Delta x_G \approx \frac{Gp}{c^3} \left[ \ln \left( \frac{2L}{x_0} \right) - 
1 \right] \, ,
\end{equation}
where $x_0$ is the initial position of the electron. Since $x_0$ must be of 
order but less than $L=cT$ and the log is a very slowly varying function we 
obtain about the same result as (\ref{yG}) for the longitudinal uncertainty,
\begin{equation}
\Delta x_G \approx \frac{Gp}{c^3} \, . \label{xG3}
\end{equation}

One should be justifiably suspicious of the Newtonian derivations since
Newtonian theory treats the gravitational field in front of the radiation
as action-at-a-distance, whereas  the gravitational field actually
propagates at $c$ and cannot  extend in front of the radiation. The nature of
the gravitational field will become clear when we discuss the calculation
using linearized general relativity.

\section{General Relativity Theory, Dimensional Estimate}

The arguments in the preceding two sections are only marginally convincing
as heuristic arguments since they are based on action-at-a-distance
Newtonian gravitational theory, with the ad hoc assumption that the energy
of the photon produces a gravitational field. In this section and the
following we give a dimensional estimate and an approximate calculation
based on general relativity theory, free of such drawbacks \cite{abs,mtw,oha}.

The field equations of general relativity are
\begin{equation}
G_{\mu \nu} = - \left( \frac{8 \pi G}{c^4} \right)T_{\mu \nu} \, . 
\label{fldeq}
\end{equation}
The left side has the units of inverse distance squared, since it is
constructed from second derivatives and squares of first derivatives of the
metric. Thus on dimensional grounds we may write the left hand side in
terms of deviations of the metric from flat, in schematic order of
magnitude dimensional form, as
\begin{equation}
LHS \approx \frac{\delta g _{\mu \nu}}{L^2} \, , \label{lhs}
\end{equation}
where $\delta g _{\mu \nu}$ denotes the deviation of the metric from 
Lorentzian, and $L$ is the same characteristic size as used in sec. III. 
Similarly the energy-momentum tensor has the units of an energy density, so 
its components must be roughly equal to the photon energy over $L^3$. Thus we 
can write the right side of the field equations schematically as
\begin{equation}
RHS \approx \left( \frac{8 \pi G}{c^4} \right) \frac{E}{L^3} \approx \frac{Gp}{
c^3 L^3} \, . \label{rhs}
\end{equation}
Equating the dimensional estimates in (\ref{lhs}) and (\ref{rhs}) we get an 
estimate for the deviation of the metric,
\begin{equation}
\delta g_{\mu \nu} \approx \frac{Gp}{c^3 L} \, . \label{dg}
\end{equation}
This deviation corresponds to a fractional uncertainty in all positions in
the region $L$, which we identify with a fractional uncertainty in position,
$\Delta x_G / L$. Thus we have an uncertainty in position due to the 
gravitational interaction given by
\begin{equation}
\frac{\Delta x_G }{L} \approx \delta g_{\mu \nu} \approx \frac{Gp}{c^3 L}\, ,
\ \ \ \Delta x_G \approx \frac{Gp}{c^3} \, . \label{xG4}
\end{equation}
As should be expected the characteristic size $L$ has canceled out of the
relation. Finally the uncertainty in momentum of the electron must be
comparable to the photon momentum, $\Delta p \approx p$, and we obtain the same
relation (\ref{xG2}) as before for $\Delta x_G$.

\section{General Relativity Theory, Approximate Calculation}

To make an approximate calculation we use linearized general relativity
theory. From the general energy momentum tensor of the electromagnetic
field it is easy to show that for radiation moving in the $x$ direction the
specific form of the energy- momentum tensor is \cite{tol,abs}
\begin{equation}
T_{mu \nu} = F_{\mu \alpha} F^{\alpha}\; _{\nu} + \frac{1}{4}F_{\alpha \beta}
F^{\alpha \beta} = \rho 
\left(  \begin{array}{rrrr}
1 \ & -1 \ \ & 0 \ \ & 0 \ \ \\
-1 \ & 1 \ \ & 0 \ \ & 0 \ \ \\
0 \ & 0 \ \ & 0 \ \ & 0 \ \ \\
0 \ & 0 \ \ & 0 \ \ & 0 \ \ 
\end{array} \right) \, , \label{emt}
\end{equation}
where $\rho=(E^2 + B^2 )/2$ is the energy density of the radiation field, and 
may be a function of $x -ct \, , \ y \, , \text{ and } z$, corresponding to a 
truncated plane wave. The equations of linearized general relativity theory 
follow from (\ref{fldeq}), with the metric taken to be Lorentz plus a small 
perturbation, $g _{\mu \nu} = \eta_{\mu \nu} + h_{\mu \nu}$.  They are
\begin{equation}
\Box \left [ h_{\mu \nu} - \frac{1}{2} \eta_{\mu \nu} h \right] = - \left( 
\frac{8 \pi G}{c^4} \right) T_{\mu \nu} \, , \ \ h \equiv \eta^{\mu \nu} h_{
\mu \nu} = h^{\alpha}_{\alpha}\, ,\label{lfldeq}
\end{equation}
$$ 
\Box = \frac{\partial^2}{\partial (ct)^2} - \frac{\partial^2}{\partial x^2}
- \frac{\partial^2}{\partial y^2} - \frac{\partial^2}{\partial z^2} \, ,  
$$
$$
\text{Lorentz gauge condition } \left[ h_{\mu}^{ \nu} - 
\frac{1}{2} \eta_{\mu}^{ \nu} h \right]_{| \nu} =0 \, . 
$$
It is straight-forward to solve this system with the energy momentum tensor
given in (\ref{emt}). We are interested only in the inhomogeneous solution, and
the system then reduces to a form involving only one unknown function,
\begin{equation}
h_{\mu \nu} = f(x-ct, y, z)
\left(  \begin{array}{rrrr}
1 \ & -1 \ \ & 0 \ \ & 0 \ \ \\
-1 \ & 1 \ \ & 0 \ \ & 0 \ \ \\
0 \ & 0 \ \ & 0 \ \ & 0 \ \ \\
0 \ & 0 \ \ & 0 \ \ & 0 \ \ 
\end{array} \right) \, , \ \ \
\Box f = - \left( \frac{8 \pi G}{c^4} \right) \rho \, . \label{h}
\end{equation}
For convenience we choose the energy density to be a product,
\begin{equation}
\rho (x-ct,y,z) = \rho_{\parallel}(x-ct) \rho_{\perp}(y,z) \, , \label{rho}
\end{equation}
and it then follows that the metric function $f$ is also a product of the
form $f(x-ct,y,z) = f_{\parallel}(x-ct) f_{\perp}(y,z)$, with
\begin{equation}
\left( \frac{\partial^2}{\partial y^2} + \frac{\partial^2}{\partial z^2} 
\right) f_{\perp}(y,z) = \left( \frac{8 \pi G}{c^4} \right) \rho_{\perp}(y,z) \
, , \ \ f_{\parallel}(x-ct) = \rho_{\parallel}(x-ct) \, . \label{wveq}
\end{equation}

Until now the energy density function has been arbitrary. We now choose a
specific function which is convenient for our purposes. For this we take a
cylinder as the envelope of the radiation, inside of which the energy
density oscillates at twice the frequency of the radiation field; the
oscillations however may be ignored since we are interested in an
average uncertainty in position. Thus we take the energy density to be a
constant $\rho_{0}$ inside a cylinder of length $L$ and radius $R$, with $R$ 
and $L$ comparable in magnitude. Then the solution to (\ref{wveq}), in 
cylindrical coordinates $x-ct\, , \ r \, , \ \varphi$, is
\begin{equation}
f = \frac{4G(E/c^2)}{L} g(r) \theta_{L}(x-ct) = \frac{4Gp}{c^3 L} g(r) 
\theta_{L}(x-ct) \, , \label{sol}
\end{equation}
$$
g(r) = \left\{ \begin{array}{lr} 
r^2/R^2 \, , & r<R \\
1+ \ln(r^2/R^2) \, , & r>R
\end{array} \right\} \, , \ \ \theta_{L}(x-ct) \equiv \theta(x-ct) \theta(ct-x
-L) 
$$
Here $E= \pi R^2 L \rho_{0}$ is the total energy of the radiation, and 
$\theta_{L}$ is the ``double'' theta function defined above - which is equal to
$0$ ahead of and behind  the radiation cylinder. Notice that the gravitational 
field does not extend ahead of the radiation field, unlike in the Newtonian 
theory. Notice also that the apparent logarithmic divergence of 
the external field is not a problem since physical effects involve the 
derivative of $f$ which falls off like $1/r$.  According to (\ref{h}) and 
(\ref{sol}) the metric takes the rather elegant form
\begin{equation}
ds^2=c^2 dt^2 - (dx^2 + dy^2 + dz^2) + f(cdt - dx)^2 \label{g}
\end{equation}

It is clear that since $g(r)$ is of order 1 inside and near the cylinder 
of radiation the deviation of the metric from Lorentz is there of order $4Gp /
c^3 L$ and we thereby obtain, using the same idea as in section V,
\begin{equation}
\frac{\Delta x_G }{L} \approx \frac{Gp}{c^3 L} \approx \frac{G \Delta p}{c^3 L}
\, \ \ x_G \approx \frac{4G \Delta p}{c^3} \, . \label{xG5}
\end{equation}

We may also estimate the gravitationally induced motion of the electron
using the above  metric and the geodesic equation of motion, to obtain an
alternative derivation of the position uncertainty. The geodesic equation
of motion is
\begin{equation}
\frac{d^2 x^i}{ds^2} + \left\{^{\ i}_{\alpha \beta} \right\}
\frac{dx^{\alpha}}{ds} \frac{dx^{\beta}}{ds}=0 \, , \ \ \alpha,\beta=0,1,2,3
\, , \ \ i=1,2,3 \, . \label{geo}
\end{equation}
For an electron moving reasonably slowly the line element is about  $ds=cdt$ 
and the $\alpha=\beta=0$ term dominates, so the equations of motion are the 
usual Newtonian limit equations
\begin{equation}
\frac{d^2 x^i}{dt^2}= - \left\{^{\ i}_{0 0} \right\} c^2 \, . \label{geo2}
\end{equation}
From the metric (\ref{g}) the Christoffel symbols are easily found and we 
obtain for motion in the longitudinal $x$ and transverse $r$ directions
\begin{equation}
\frac{d^2 x}{dt^2} = \frac{1}{2} \frac{\partial f}{\partial x} c^2 \, , \ \
\frac{d^2 r}{dt^2} = - \frac{1}{2} \frac{\partial f}{\partial r} c^2 \, .
\label{geo3}
\end{equation}
where we assume motion in the $x$ and $r$ direction.

Longitudinal motion is easy to analyze and rather informative. The
derivative of $f$ gives two delta functions. As the front of the radiation
cylinder passes the electron it first gives the electron a velocity impulse
of
\begin{equation}
\Delta \dot{x} = \frac{2Gp}{c^2 L} g(r) \, , \label{impx}
\end{equation}
and then as the back of the radiation cylinder passes the electron receives
an equal and opposite velocity impulse and stops. In the time of passage it
has moved
\begin{equation}
\Delta x_G = \frac{2Gp}{c^2 L} g(r) T = \frac{2Gp}{c^3 } g(r) \approx
\frac{2Gp}{c^3 } \, , \label{xG6}
\end{equation}
since $g(r)$ is of order 1 in and near the radiation cylinder. Thus we obtain
the same result (\ref{xG2}) as previously. Notice that the gravitational field 
of the radiation only acts as it passes over the electron, not before and not
after.

For the transverse motion we differentiate $f$ with respect to $r$ to find
\begin{equation}
\frac{d^2 r}{dt^2} = -\frac{4Gp}{c^2 L}
\left\{ \begin{array}{lr} 
r/R^2 \, , & r<R \\
1/r \, , & r>R
\end{array} \right\} \theta_{L}(x-ct)
\end{equation}
so that in the region of the cylinder we have very roughly a velocity
impulse and a corresponding position change $\Delta r \approx (\Delta \dot{r}
/2) T$, given by
\begin{equation}
\Delta \dot{r} = \frac{4Gp}{c^2 LR}T \, , \ \ \Delta r= \frac{2Gp}{c^2 LR}T^2
\approx \frac{2Gp}{c^3}
\end{equation}
That is once again we find that the transverse motion corresponding to an
uncertainty in position due to gravity is the same as in (\ref{xG}) or 
(\ref{xG2}).

\section{The Minimal Distance Uncertainty}

The GUP has a remarkable consequence. If the photon momentum and $\Delta p$ are
chosen to be very small then the electron position is imprecise because the
long photon wavelength gives poor resolution. If the photon momentum and 
$\Delta p$ are chosen to be very large, then the gravitational field of the 
photon makes the electron position very imprecise. Between the two extremes 
there is a minimum position uncertainty, which we find from (\ref{gup}) to be
\begin{equation}
\Delta x_{\text{min}} \approx 2\sqrt{\frac{G \hbar}{c^3}} = 2 L_p \, , \
\text{ for } \ \Delta p \approx \sqrt{\frac{\hbar c^3}{G}}\frac{E_p}{c} \,.
\end{equation}
This means that we can never localize the position of a particle such as an
electron to better than about the Planck distance.

Similar analyses of spacetime using the path integral formalism lead to
analogous conclusions; spacetime at small distances and times undergoes
quantum fluctuations, and at the Planck scale the fluctuations are of the
same order as the distances involved \cite{mtw,amel}.
	
\section{conclusions}

We have shown, using Newtonian and general relativistic gravity, that the
position-momentum uncertainty principle of quantum mechanics is modified by
an additional term. In both theories it is clear that the extra term must
be proportional to the energy or momentum of the photon, so on purely
dimensional grounds the order of magnitude of the extra term is uniquely
determined. As a consequence there is an absolute minimum uncertainty in
the position of any particle such as an electron. Not surprisingly the
minimum is of order of the Planck distance.

In view of the absolute minimum position uncertainty one may plausibly
question whether any theory based on shorter distances, such as a spacetime
continuum, really makes sense. Indeed in light of the fact that laboratory
experiments which probe small distance properties of particles are all high
energy scattering experiments, one might conclude that spacetime at such
small scales may not be a useful concept, and that spacetime at the Planck
scale may not even exist in any meaningful operational sense. Such ideas
are not new and were espoused in the era of S-matrix theory in the 1960s,
that is that the scattering amplitude expressed n terms of input and output
momenta may be the fundamental reality of high energy physics, and not
point-like or string-like particles in a spacetime continuum \cite{sm}.
One might even speculate that the spacetime continuum concept actually
impedes physics in the same way that the concept of an ether impeded
physics in the 19th century. As such, a theoretical structure based entirely
on momenta, such as a modern version of S-matrix theory, might be desirable
and interesting.

\section*{Acknowledgements}

This work was supported by NASA grant NAS 8-39225 to Gravity Probe B. John
Schwarz posed the question studied here in a talk at the Stanford Linear
Accelerator Center in fall 1996, but is of course not responsible for our
interpretation of it. Finally the Gravity Probe B theory group at Stanford
provided many stimulating discussions.

\end{document}